\PassOptionsToPackage{numbers,sort&compress}{natbib}

\documentclass[12pt]{extarticle}

\usepackage{johd}
\usepackage{amsmath}

\setcitestyle{super}

\makeatletter
\renewcommand\@biblabel[1]{[#1]}
\makeatother

\title{An Active Soft Condensed Matter Approach to the Physics of Living Systems}

\author{Nitin Kumar$^{a}$$^{*}$ \\
        \small $^{a}$Department of Physics, Indian Institute of Technology Bombay, Powai, Mumbai - 400076\\
        \small $^{*}$Email: \tt{nkumar@iitb.ac.in} \\
}
\date{}

\begin{document}
\maketitle


\begin{abstract} 
\noindent This article aims to introduce the broad field of soft active matter physics and its relevance to the life sciences in simple, accessible language. Although this area of research is relatively new, it has already demonstrated significant potential in providing a physical understanding of many biological processes. While several review articles by leading researchers exist, they can be difficult to grasp for undergraduate students and even early-career researchers who wish to enter this field. In this article, I cover the basics, introduce the origins of soft active matter physics, and explain how it differs from traditional equilibrium condensed matter ideas at the fundamental level. For the most part, I will avoid mathematical equations and excessive technical precision in several statements. Instead, I will focus on communicating the core ideas and the overall spirit of the argument, using everyday examples to develop a physical intuition. The primary focus will be on the dynamical aspects of these systems. I will conclude by briefly discussing a published experimental study from our research group that examines universal features of the trajectories of homing and migrating organisms.\\ \end{abstract}

\noindent\\




\section{Overview}
If I ask you to imagine living matter with your eyes closed, you would likely picture cells dividing, bacteria swimming, birds migrating, insects crawling, or animals hunting. These systems typically take the centre stage in the domain of biology. Now think about your Physics courses. The objects that you often encounter are electrons, atoms, molecules, blocks sliding down inclined planes, masses hung to springs or strings, pulleys, and so on. What is common in all these examples? They are all non-living, unlike objects in Biology. In our standard physics courses, we rarely discuss living systems.  While we apply Newton's laws to calculate the trajectories of a non-living object of mass $M$, we hesitate to do the same with a living system of the same mass. Why can you not use the same equations that predict the motion of a ``dead" projectile of mass $M$ to a live object of the same mass? Is it possible to conceptualize a ``living matter" built from a collection of ``living atoms" or ``living molecules"? And in the process of doing so, can we formulate a robust physical framework to understand you and me, or birds, animals, microorganisms, etc.?

In this article, I will introduce you to a novel physical framework that has emerged as a promising paradigm to understand the Physics of living systems. It is called \textit{Active Matter Physics}, which is actually a shorthand for the rather wordy term \textit{Active Soft Condensed Matter Physics}. I will dissect this title word by word and introduce this framework in simple, straightforward language. Along the way, I will explain the fields of condensed matter, soft matter, and active matter, shedding light on their deep interconnection. Finally, I will provide an example from our own research work and highlight some universal features that encompass the motion of living organisms. 

\section{What is Condensed Matter?}
Let’s start with the basics: Matter. In the simplest terms, matter is a collection of atoms or molecules. This collection exists in a state where its physical properties are emergent, that is, they arise from collective behavior and are qualitatively different from the properties of the individual atoms. We all know that such a collection exists in three primary phases: Solid, liquid, and gas. To understand these phases physically, we look at the total energy of the system. Energy is an abstract quantity, conserved in total, but split into two parts: kinetic energy (\textit{KE}) and potential energy (\textit{PE}). Thermal \textit{KE} originates from the random thermal motion of atoms
at temperature $T>0$. \textit{PE} is the energy of an attractive interaction pulling atoms together to an equilibrium separation of $r_0$ [Fig. 1(a)]. 

We can visualize the phases of matter as the outcome of a competition between these two forms of energy.
\begin{itemize}
    \item \textbf{Gas:} If \textit{KE} $>$ \textit{PE}, the particles fly apart and move independently.
\end{itemize}

\begin{itemize}
    \item \textbf{Solid:} If \textit{KE} $<$ \textit{PE}, the attractive forces win and atoms lock into place, typically in a space-periodic manner, to minimize the total energy. Their mean separation $r_0$ corresponds to the minimum of the potential energy plot in Fig. 1a.
\end{itemize}

\begin{itemize}
    \item \textbf{Liquid:} If \textit{KE} $\approx$ \textit{PE}, i.e., energies are comparable. But it is not that simple. Strictly speaking, we must also confine them, without which they will, over a large timescale, either disperse away from each other. This requirement introduces the concept of pressure. Therefore, a liquid phase emerges when a collection of atoms is confined, and the effects of \textit{KE} and \textit{PE} effectively cancel each other, leading to a dense state, with atoms still separated by a distance approximately equal to $r_0$, without permanent neighbors. 
\end{itemize}

Out of these phases, condensed matter refers to the ones where the atoms stay in proximity to each other. That means only solids and liquids belong to the category of condensed matter.

\section{What is soft in soft condensed matter?}
Let us try to understand solids and liquids from a mechanical perspective. Upon touch, solids (such as iron blocks and ice cubes) feel very hard to change their shape, whereas liquids are ``self-deforming", meaning we can change their shape without any effort. But some objects feel neither purely solid nor purely liquid upon touch. Like toothpaste, shampoo, clay, foam, gels, and food items like dough, among others. These materials are classified as \textit{Soft Matter}, characterized by their ease of deformation. 

Let us now define the term \textit{deformation}. It simply means a change in shape. You can change the shape of an object either by changing its volume, known as compression, or without changing the volume, known as shear. It is tough to compress either solids or liquids because this requires bringing atoms much closer than $r_0$, forcing them to climb up the repulsive part of the potential energy curve shown in Fig. 1a. Therefore, both solids and liquids are considered nearly incompressible. However, they differ in terms of shearability. 

\begin{figure*}[t!]
	\centerline{\includegraphics[width= 0.9\textwidth]{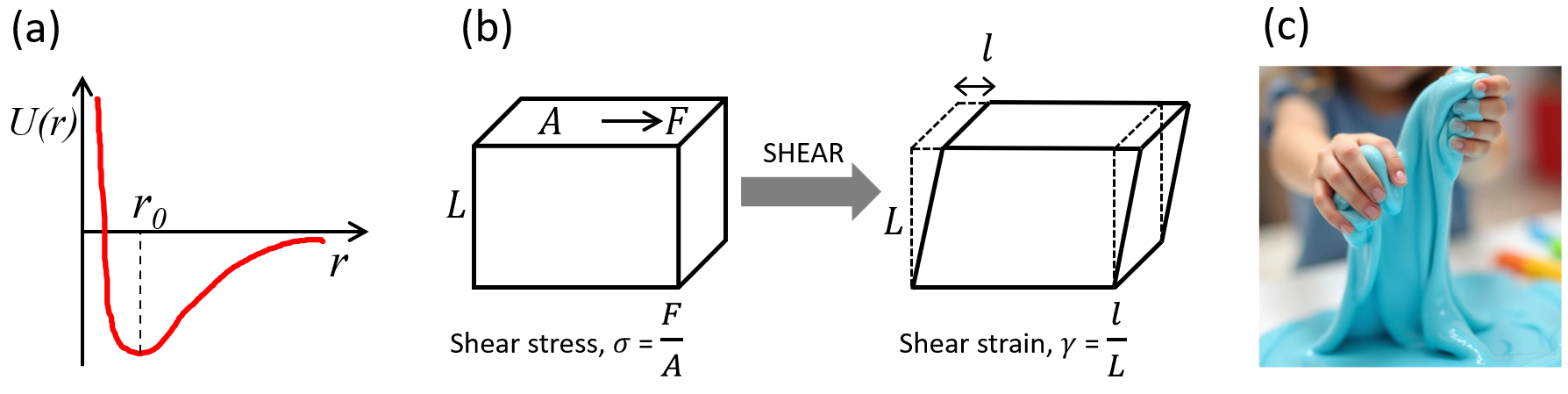}}
	\caption{(a) A typical interaction energy curve signifying the attractive and repulsive regions for $r>r_0$ and $r<r_0$, respectively. (b) A schematic of shear deformation of a material. (c) Slime, a typical soft matter system.}
	\label{fig7}
\end{figure*}

We can quantify shearability using the shear modulus, $G$. From Fig. 1(b), it is defined as the ratio of shear stress ($\sigma = F/A$) to shear strain ($\gamma = l/L$), i.e., $G = \sigma/\gamma$. 

Let us obtain a rough estimate of $G$ for typical materials. We can estimate $G$ using dimensional analysis as energy per unit volume ($G = E/r_0^3$).

\begin{itemize}
    \item \textbf{For a Metal (Solid):} The energy holding the atoms together is the metallic or ionic bond energy, ranging $1 - 10 \text{ eV}$ (or $40-400$ $k_BT$ at $T$ = 300 K, where $k_B$ is the Boltzmann constant and $k_BT$ corresponds to the energy associated with temperature T). For our calculation, let us use $E$ = 5 \text{eV}.  The length scale is the atomic bond length ($ r_0\approx  $ 2.5 $\text{\AA}$). The math works out to $G = 5.1 \times 10^{10}$ $\text{Pa}$ = $51$ $\text{GPa}$.

This is massive. It would take an immense amount of force to shear a metal block.
\end{itemize}

\begin{itemize}
    \item \textbf{For water (Liquid):}  There are no permanent bonds between water molecules that lock them into fixed lattice locations like a solid. Their thermal kinetic energy balances the interatomic attraction. Thus $E$ = 0. That is why if you tilt your water bottle, it does not resist changing shape and shears effortlessly.  Therefore, $G$ = 0 for liquids. 
\end{itemize}

\begin{itemize}
    \item \textbf{For Soft Matter:} That's where things get interesting. Let us estimate $G$ for, say, toothpaste. But first, we need to know what it is made of. It is not a simple atomic lattice of atoms. Rather, it is composed of macromolecules, polymers, and colloids. These entities are much larger than a single atom, typically lying in the range of a micron to a tens of nanometers range ($10^{-6} - 10^{-8}$ m). Moreover, the interactions between them are not strong covalent or metallic bonds; they are typically weak, like hydrogen bonds and Van der Waals forces. Thus, the energy scale is much lower, often of the order of the thermal energy $1-20$ $k_B T$ at room temperature, which is equivalent to $0.025-0.5$ \text{eV}. For our calculation, let us use $E = 0.05$ eV and $r_0 = 10^{-8}$ m. Therefore, for toothpaste $G$ $\approx$ 8000 Pa, which is smaller than that of a solid, but more importantly, not zero like liquids.    
\end{itemize}

This gives us our definition: Soft Matter includes materials with a shear modulus between that of a solid (very high) and a liquid (zero). They deform easily because their internal structural bonds are weak, and individual constituents are much larger than the atomic scale. In practice, most familiar examples, such as clay, gels, foams, and the food we consume, typically have shear moduli in the range $10^2$ to $10^5$ $Pa$.

 The following table and a one-dimensional axis of $G$ summarize the difference between solids, liquids, and soft matter.

 \begin{table}[h!]
\centering
    \begin{tabular}{|l|c|r|c|}
\hline
 & Solid & Liquid & Soft Matter \\ \hline
Bond energy, $E$ & $1 - 10$ \text{eV}  & 0 & $0.02 - 0.5$ \text{eV} \\
Bond length, $r_0$ & $1-5$ $\text{\AA}$  & $1-5$ $\text{\AA}$ & $10^{-6} - 10^{-8}$ m \\
Typical Shear Modulus, $G$ (Pa) & $5.1 \times 10^{10}$ $\text{Pa}$ & 0 & 8000 $\text{Pa}$ \\ \hline
\end{tabular}
\caption{The table summarizes the typical energy and
length scales, along with the corresponding shear moduli,
for solids, liquids, and soft materials.}
\end{table}

\begin{center}
	\includegraphics[width=16.83cm, height=2.65cm]{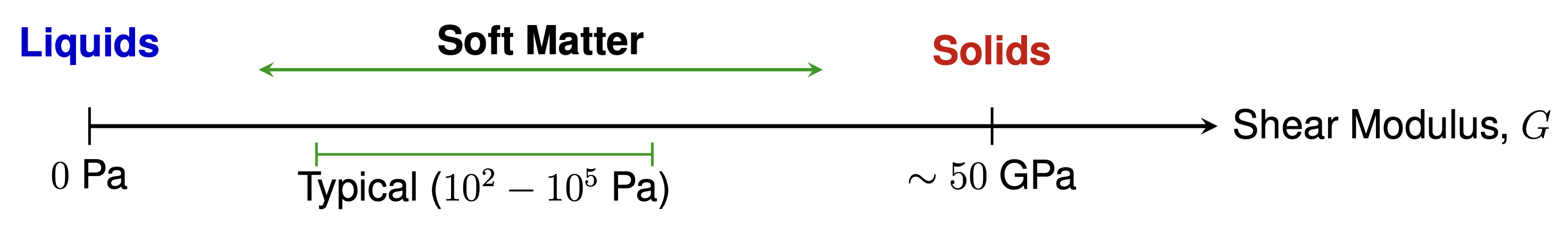}
\end{center}

Low values of $G$ for soft materials introduce their own set of challenges and make their physical behavior difficult to predict. Weak forces, such as gravitational and thermal forces, can cause their constituents to shift from their equilibrium positions, and this too over time scales of seconds to minutes. As a result, formulating a simple and predictive physical description becomes a challenging task. However, since they deform over time and energy scales that humans can easily comprehend, they become extremely appealing and fun to play with. This might explain why kids are fascinated with objects like clay, slime, and jelly [Fig. 1c].


\section{Active particles and active matter: The Kinematic View}
This is where we transition from standard thermodynamics to non-equilibrium physics. A few technical remarks are necessary at this stage. Readers primarily interested in the conceptual picture may skip the footnotes, which elaborate on the underlying mechanisms.

Let's begin. Imagine looking under a microscope.

\begin{itemize}
\item\textbf{Scenario A:} You see a pollen grain ($\approx$ 5 $\mu$m in size) jiggling in water. It is not because the pollen grain is alive. This motion is merely a manifestation of thermal energy. Let me explain. It moves due to random collisions of thermally agitated water molecules from all directions over time, a phenomenon known as Brownian motion. The word used for such time-random processes is \textit{stochastic}. Since the water's temperature remains constant, the system is at thermal equilibrium. Consequently, the probabilities of states of energy $E$ are governed by the Boltzmann weight $e^{-(E/k_B T)}$. Since the kinetic energy of the particle is $E = (1/2) mv^2$, its instantaneous velocity distribution follows the Maxwell–Boltzmann form: $P(v)\propto e^{-(mv^2/2k_B T)}$. Moreover, if we track every water molecule colliding with the grain at each instance of time, we can, in principle, apply Newton's equation of motion and predict the motion of the pollen grain precisely\footnote{These dynamics follow another important principle in equilibrium thermodynamics: \textit{detailed balance}. When the particle transitions from velocities $v_1$ to $v_2$ and vice versa, with transition rates $W(1\rightarrow 2)$ and $W(2\rightarrow 1)$ respectively, they follow the equation $W(1\rightarrow 2) P(v_1) = W(2\rightarrow 1) P(v_2)$. This ensures there is no net probability current in the system.}. Since Newton's equation involves the second derivative of time, the dynamics remain invariant under time-reversal, i.e., if $t \rightarrow -t$. In practical terms, this implies that a movie of a jiggling pollen grain, played forward or backward, would look statistically indistinguishable or time-symmetric\footnote{In more technical terms, the probability of observing time-forward and time-reversed phase-space trajectories (i.e., trajectories involving both position and velocity: $(\mathbf{r}(t), \mathbf{v}(t)$ and $(\mathbf{r}(-t), \mathbf{-v}(-t))$ over a time-scale $t$, denoted by $P(t)$ and $P(-t)$ respectively) will be equal. The net entropy production rate over a time $t$ is given by $(1/t) \ln [P(t)/P(-t)]$, which is clearly zero in this case as $P(t)=P(-t)$. This equation originates from a class of mathematical results known as Fluctuation Relations.}. Such a system is said to be in a thermal equilibrium state, and particles displaying these dynamics are called \textit{passive particles}. Atoms and molecules that make traditional phases of matter fall into this category.

\item \textbf{Scenario B:} You see a bacterium, which is also a microscopic particle but alive in a true sense. However, it does not just jiggle; it also propels in a specific direction\footnote{It is possible to encounter situations in other species of microbes that exhibit directed rotation rather than directed translation. These are called active chiral particles. Even more complex behaviors can arise when directed motion is absent, but jiggling around mean positions violates the equipartition theorem, leading to unequal kinetic energy in different directions. All of these are equally valid examples of active particles, as they break time-reversal symmetry. To keep the language simple, I will continue to use the term directed motion for all such cases.}. It achieves this motion by converting stored internal energy, generated by the metabolism of the food it consumes, into mechanical energy. Now, recall Carnot's engine from thermodynamics: it extracts heat from a hot reservoir, performs work, and dissipates the remaining heat into a cold reservoir, all in accordance with the first law of thermodynamics. By analogy, the bacterium moves in a directed manner because it acts like a chemical engine operating irreversibly between two chemical potential reservoirs. Clearly, this is not a state of thermal equilibrium as there is a net entropy production\footnote{At a microscopic level, the time-reversal symmetry is broken, and the probability of observing time-reversed phase-space motion [$P(-t)$, in comparison to $P(t)$] becomes increasingly rare as $t$ grows.}. Consequently, one can easily distinguish between forward and reverse movies of the bacterium swimming. Moreover, the direction in which the bacterium propels becomes effectively unpredictable. It ``decides" its own direction by modifying the beating pattern of its flagella, for example\footnote{While the motion arises from underlying biochemical processes, the coupling of many internal degrees of freedom and intrinsic stochasticity makes the swimming direction unpredictable in practice.}. Such objects that are driven out of their thermodynamic equilibrium, without an explicit presence of an externally applied field, are known as \textit{active particles}. Their dynamics are governed by their internal state, rather than the principle of equilibrium statistical mechanics, as was the case with a passive particle. These active particles are what I referred to as ``living atoms" and ``living molecules" in the introduction.

\end{itemize}

To emphasize this contrast further, a passive particle can move in a directed fashion only in the presence of an external force or gradient. For example, a charged particle placed in an electric field acquires a drift velocity determined by that field. Active particles, in contrast, always exhibit self-directed motion. A bacterium, for instance, propels itself using internal biochemical energy derived from food, and the food does not command the direction of its motion\footnote{It is important to highlight a distinction. In various biological settings, the motion can indeed arise due to external gradients like chemotaxis, phototaxis etc. However, these fields do not produce a force which is equal to the gradient of a conservative potential, as is the case with the electric field acting on an electron, for example. The motion of a living particle remains self-propelled, albeit ``influenced", rather than propelled, by such fields.}. The same applies to us. The food we consume does not decide in which direction and how fast we move.

To conclude, traditional, non-living matter, at equilibrium, is composed of passive particles, such as atoms and molecules. In contrast, active matter comprises active particles, a definition that encompasses all living systems across scales from subcellular components and individual cells to tissues, organs, and even entire organisms. Therefore, the active matter framework provides a natural starting point for building a physical framework that governs all biological systems.

\begin{center}
	\includegraphics[width=10cm, height=5.333cm]{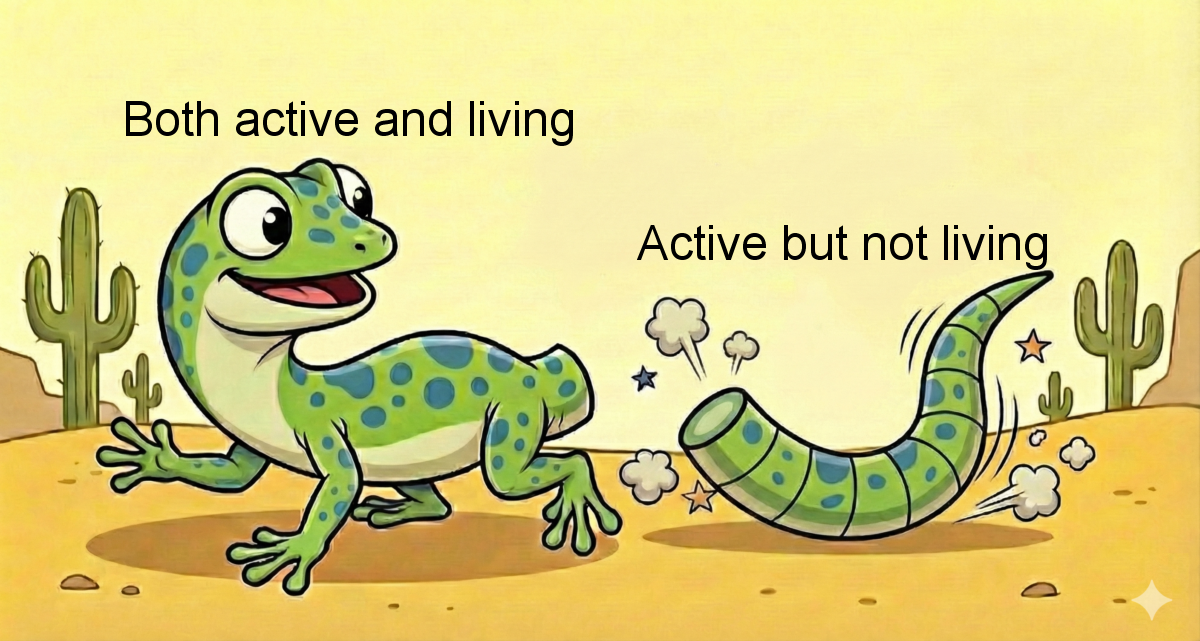}
\end{center}

\section{A subtle difference between living and active matter}
\textit{Living matter} refers to systems that are literally alive and composed of smaller constituents, which may also be living. For example, all organisms are living matter. On the contrary, active matter and active particles themselves are not living, yet are capable of self-driven motion that resembles living matter. Examples include cars, drones, toys, robots, molecular motors inside our cells, etc. Experimentalists have developed a variety of artificial active systems, including catalytic colloidal particles, vibrated granular particles, programmable robots, and in-vitro biopolymer-motor protein mixtures, to name a few. They all harness ambient or stored energy sources to undergo ``life-like" motion. 

To illustrate this distinction, I will provide a silly yet simple example. Consider a gecko shedding its tail as a defense mechanism. When the tail is detached, it continues to shake and wiggle. This motion is due to muscle contractions powered by the hydrolysis of adenosine triphosphate (ATP) to adenosine diphosphate (ADP), a chemical reaction that stores energy. The wiggling tail distracts the predator, creating confusion about which part is the real animal. In this situation, the tail is no longer alive, but from a dynamical perspective, it remains active. It moves autonomously by converting its internal energy. This motion will stop once all the ATP is consumed. Meanwhile, the gecko itself is both alive and active, and it moves with purpose as it escapes. Thus, from the perspective of motion alone, both the detached tail and the gecko are examples of active systems. However, at the organism
level and in a sustainable manner, only the gecko is living.

Thus, the following sentence summarizes the difference between living and active matter: \textit{all living matter is active, but not all active matter is living}.

\section{Why is all living matter also soft matter?}

Or, in other words, why is living and active matter considered a sub-branch of soft matter physics? Let me illustrate this using our own bodies as an example of active materials. 

Before that, we must identify fundamental length and energy scales involved. An amino acid can be considered the fundamental building block of the human body's solid mass. They are typically $4-8$ $\text{\AA}$ in size and consist of elements like carbon, hydrogen, oxygen, and nitrogen, etc. These are connected together by strong covalent bonds with energy scales of the same order as metallic bonds (1-10 \text{eV} or $40-400$ $k_BT$ at room temperature). On the other hand, the fundamental energy scale typically arises from the hydrolysis of ATP to ADP. It typically releases energy of the scale $0.3-0.6$ $\text{eV}$ ($10-20$ $k_BT$). This energy scale is much smaller than the energy required to break covalent bonds over amino acid length scales. For our bodies to be classified as active matter, there must exist structures of sizes whose interaction strength is of the same order as the energy released during ATP hydrolysis, roughly equivalent to hydrogen bonds or van der Waals forces. Only then will the condition of local energy-injection be met, which is necessary to term a material active.

It turns out that various biological structures across different length scales, ranging from subcellular components such as the nucleus, mitochondria, and cytoskeletal elements to tissue-level structures such as skeletal muscle and collagen fibers, meet the above criteria. They typically range in size from 10 nm to 10 $\mu$m and interact with weak forces. At this length scale, the biological matter is
soft as it can undergo easy shear deformation and reorganization, thereby acquiring activity. That makes our bodies a classic example of soft living active matter. The same also applies to artificial active matter systems, which are all easily deformable.

In contrast, consider a block of iron. To shear deform it or drive it out of equilibrium at the microscopic level, one would need a mechanism capable of supplying energies on the order of hundreds of $k_B T$, or several electron volts, corresponding to the strength of metallic bonds. While this is possible in theory, such mechanisms do not occur naturally in an autonomous and sustained manner. Moreover, it lacks larger macrostructures that can undergo local deformations by utilizing ambient energy sources. Thus, a block of iron is neither soft nor active.

Therefore, most systems that exhibit spontaneous activity in nature are also soft. In this sense, we are soft as we are easily deformable, but we are also living and active.

\begin{center}
	\includegraphics[width=11.7cm, height=6.4cm]{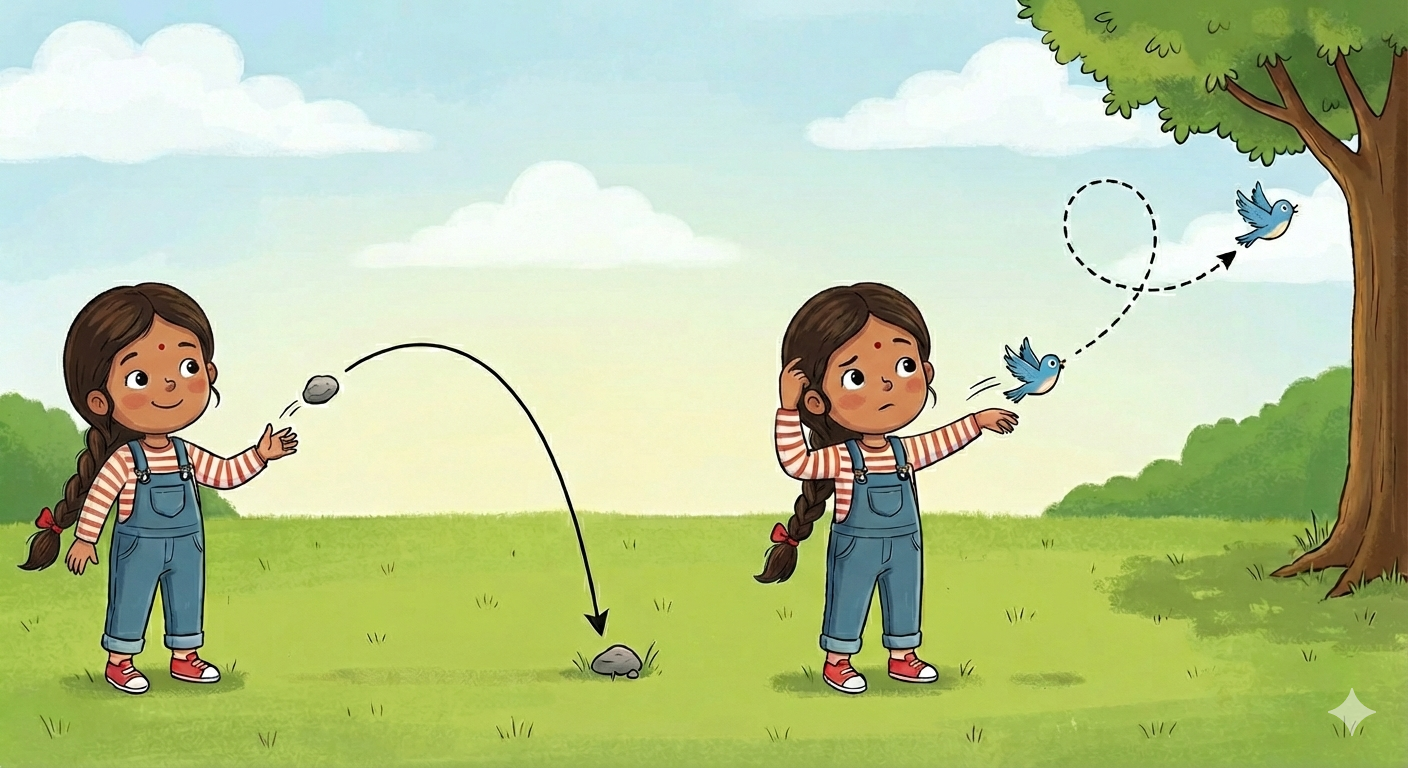}
\end{center}

\section{Why is it difficult to describe the mechanics of active particles and active matter?}

Here I will answer the question I posed at the beginning: Why can Newton’s laws predict the motion of a dead projectile of certain mass but fail to describe the dynamics of a living system of the same mass, even though both are physical objects?

In classical mechanics, we use Newton's equation of motion to predict a particle's trajectory, ($\mathbf{r}(t)$, $\mathbf{v}(t)$). However, you can do so only when working in a non-accelerating frame of reference, known as inertial frames. If you attempt to blindly apply Newton's law ($\mathbf{F}_{ext} = m\mathbf{a}$, where $\mathbf{F}_{ext}$ is the applied force) to predict the trajectory of a projectile in a non-inertial or accelerating frame, the trajectory will not follow the predicted path. The projectile will deviate from the anticipated trajectory, and you might end up wondering that the projectile has a ``mind of its own". However, if you incorporate the pseudo forces, $\mathbf{F}_{p}$, that originate from the acceleration of the reference frame (i.e. $\mathbf{F}_{ext} + \mathbf{F}_{p} = m\mathbf{a}$ where $\mathbf{F}_{p}$ denotes centrifugal or Coriolis forces etc.), you can predict the trajectory precisely. 

Now, let us return to an active particle in an inertial frame of reference. Take, for example, a bird of mass $M$. If we toss a bird up in the air, it will fly away in an unpredictable direction, and its trajectory will not follow Newton's equations applied to a mass $M$. In this case, the bird indeed has a mind of its own, and the deviation of the trajectory is not due to any pseudo force as we continue to remain in an inertial frame of reference. In this case, we have to incorporate time-dependent forces originating inside the bird from the conversion of internal energy into self-propulsion. If we denote such forces by $\mathbf{F}_{active}(t)$, the Newton's equation gets modified to $\mathbf{F}_{ext} + \mathbf{F}_{active}(t) = M\mathbf{a}$. Therefore, even though Newton’s laws remain fundamentally valid, such systems generate a time-dependent force $\mathbf{F}_{active}(t)$ whose source is the complex biology unique to the individual organism and becomes practically impossible to predict. As a result, unlike passive projectiles in inertial frames, their trajectories cannot be predicted without explicitly modeling their internal dynamics and energy consumption in the form of $\mathbf{F}_{active}(t)$. That is why, for living matter, their biology determines their future position and velocity, rather than $\mathbf{F}_{ext}(t)$ as for passive particles in the Newtonian formalism. Therefore, since deterministic prediction from initial position and velocity alone is not feasible, the aim of the active matter framework is to develop statistical or coarse-grained laws that capture universal features of such motion, as we will see in the next section.

 \begin{center}
	\includegraphics[width=12.5cm, height=4.375cm]{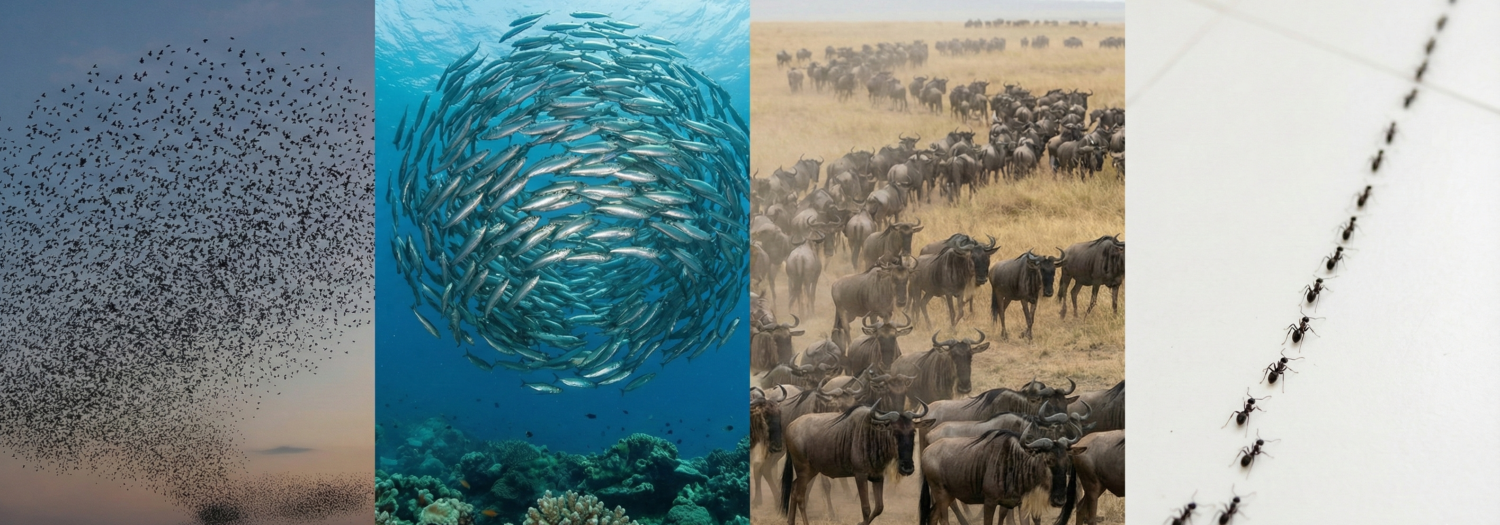}
\end{center}

This difference persists even at the collective level at the active matter scale. For traditional condensed matter composed of passive particles, an isolated system retains its total energy. This provides a natural starting point for equilibrium statistical mechanics. By setting the condition that thermodynamic equilibrium corresponds to the state of maximum entropy, state variables like temperature ($T$), pressure ($P$), and chemical potential ($\mu$) can be defined. Moreover, these state variables are related to each other through equations of state (for example, the ideal gas equation $PV = Nk_BT$ for $N$ particles).

Active and living matter, however, operates in a fundamentally different regime. Each active particle continuously consumes energy and injects it locally into the system, typically by operating between high- and low-chemical-potential reservoirs, as explained previously. As a result, the system reaches a non-equilibrium steady state characterized by sustained energy flux and finite entropy production rate. In such systems, familiar equilibrium concepts and state variables like $P$, $T$, etc. do not automatically apply. Consequently, at the collective level, active matter exhibits striking behaviors with no direct equilibrium counterpart. There are countless examples of physical phenomena that have no direct counterpart in equilibrium passive matter.

This explains why images of flocking birds, schooling fish, or collections of bacteria hold a special place in the active matter community, as such collective behaviour in their dynamics and structure challenge our intuition built on equilibrium physics. Formulating a universal framework as powerful as equilibrium statistical mechanics for active and living matter remains an open challenge and a playground for novel physical discoveries.

\section{Homing and migratory paths of living organisms}

Here, I will transition from pedagogical discussion to a specific research problem conducted in our laboratory, in collaboration with a research group at IIT Mandi. One of the key takeaways of this article so far has been that living systems follow unpredictable trajectories because their paths are ultimately determined by their biological behavior rather than an externally applied force. But can we still extract meaningful insights from their motion? And in the process, can we predict or even quantify certain behavioral traits? This question inspired one of our recently published research projects \cite{PRXLife}, in which we attempted to uncover universal features of trajectories traced by living organisms.

Our motivation was as follows. Living organisms not only exhibit active motion but, in many cases, also have a well-defined destination. Numerous examples from biology and ecology illustrate this behavior. Animals often forage over large distances in search of food and subsequently return to their home or territory. This behavior is known as homing, which is achieved by repeatedly navigating through unfamiliar locations. Migratory birds travel vast distances across unknown landscapes between their breeding and feeding grounds. This broad definition also applies to microorganisms undergoing chemotaxis and phototaxis. One such trajectory is illustrated in Fig. 2b, which shows the homing paths of pigeons of varying flock sizes, covering a distance of approximately 8 km. The question is: is there anything to learn from such trajectories, and are there universal features in these paths?

\begin{figure*}[t!]
	\centerline{\includegraphics[width= .85\textwidth]{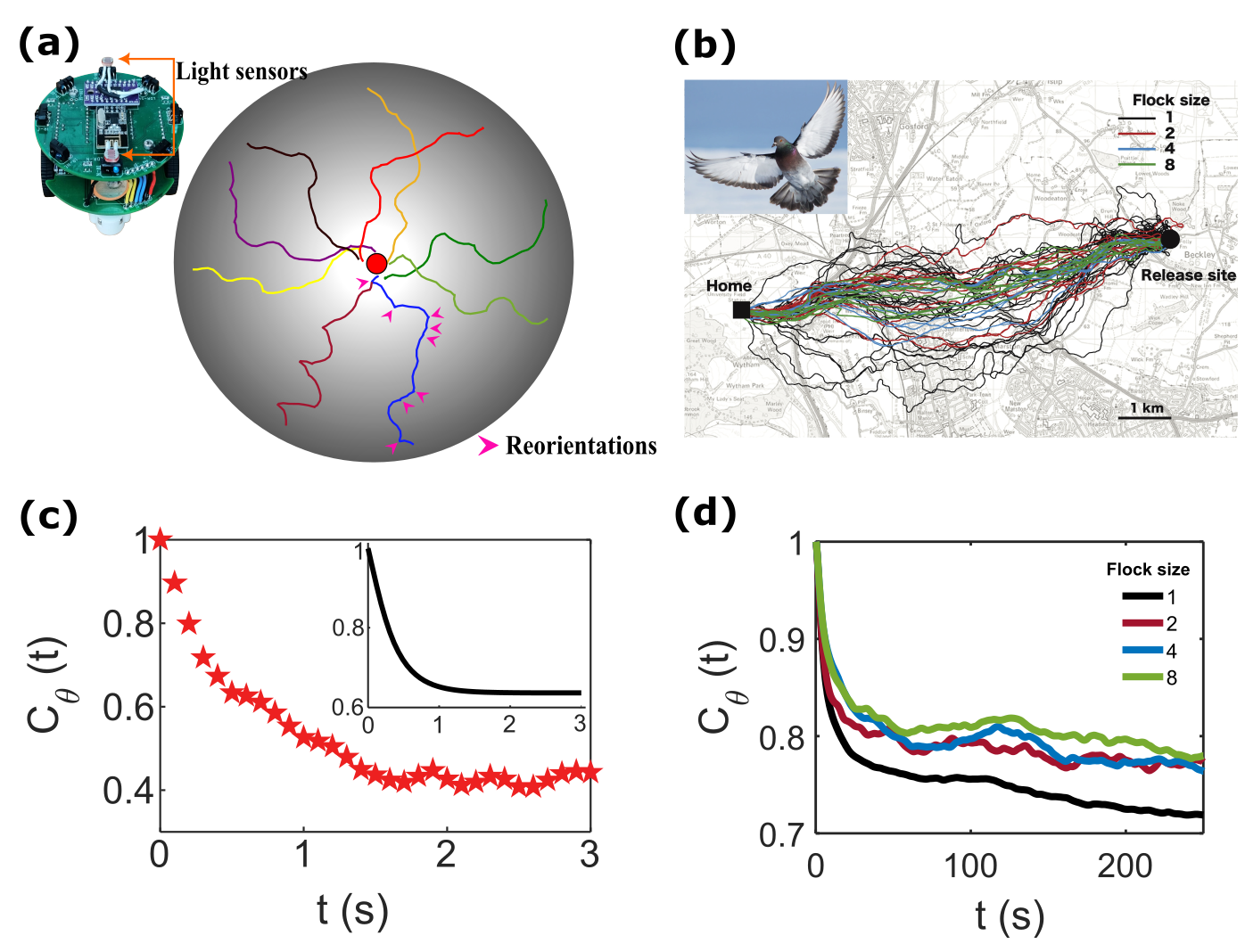}}
	\caption{(a) The robot used in experiments has two light sensors at its front and back. It performs a homing from the perimeter to the centre of a circular region, lit by a radially symmetric light-intensity gradient field. Typical homing trajectories are represented in different colors, with magenta symbols highlighting the locations of reorientations shown for one typical trajectory. (b) Trajectories of homing pigeons of different flock sizes. (c) The temporal orientational autocorrelation function shows a decay and finite saturation, qualitatively matching the theoretical expression shown in the inset. (d) Similar correlation trends are also observed in homing pigeons.}
	\label{fig7}
\end{figure*}

To gain a deeper understanding of this phenomenon, we break down this motion into two parts. 
\begin{itemize}
\item \textbf{Inherent Stochasticity}: In principle, the most efficient path for an animal to migrate would be a straight line connecting its starting point to its destination. In reality, animals traveling over vast distances rarely follow such a direct route. This is due to many factors, including favourable or adverse weather conditions, geographical barriers, predator avoidance, and temporary detours in search of food or water. These factors naturally introduce an inherent stochasticity into their motion.
\end{itemize}

\begin{itemize}
\item \textbf{Reorientations and course corrections}: 
During migration or homing, animals actively navigate toward their destination, continually adjusting their heading as they move. They rely on a variety of mechanisms, including sensing the Earth’s magnetic field, using topographical and celestial cues, or following olfactory signals. As a result, their stochastic trajectories are continuously guided by decision-making and course-correcting reorientations. 
\end{itemize}

The final paths, therefore, must emerge from a competition between random fluctuations and directed reorientation. With this insight, we set out to investigate whether universal statistical features exist in homing and migratory trajectories of living systems.

In our laboratory, we study this problem using programmable robots as model active systems \cite{EPJE}. These robots, 7.5 cm in diameter (Fig. 2a inset), are active, as they use chemical energy stored in batteries to move. To model the inherent stochasticity in their dynamics, we program them to experience a constant-magnitude stochastic force that is randomly varying in direction, causing them to exhibit Active Brownian (AB) motion. The degree of randomness in their motion is quantified by the rotational diffusion constant, $D_r$, which we can tune experimentally. Additionally, we equip the robots with light sensors that enable them to detect changes in light intensity, similar to how animals use sensory inputs to navigate their environment. These sensors are used to model animal navigational capabilities, enabling the robots to perform course-correcting reorientations toward the intended direction.

The experiments are conducted in a circular arena with a radius of 50 cm, over which we shine a radially symmetric light gradient with the highest intensity at the center, designated as the home location, and the lowest intensity at the boundary (Fig. 2a). We impose a simple rule: during homing, whenever the robot’s intrinsic randomness causes it to turn away from home, that is, toward regions of lower light intensity, it must reorient toward the center and resume its homing motion. Using this rule, we generated hundreds of homing trajectories. A few typical trajectories are shown in Fig. 2a, with reorientation events indicated with pink symbols on one of them.

To quantify the shape of these homing paths, we computed the temporal velocity autocorrelation function $C_\theta(t)=\left\langle \cos[\theta(t)-\theta(0)] \right\rangle$. It simply quantifies the probability of the normalized velocity vector at time $t$ to be equal to that at time $0$ and averaged over multiple trajectories. For example, for a particle moving in a straight line, $\theta(t)-\theta(0)$ = 0 and $C_\theta(t) = 1$ for all $t$. Fig. 2c shows this function averaged over hundreds of experimental trajectories. Inspired by these experiments, we also developed a theoretical model that generated the following expression for the $C_\theta(t)$:

    \begin{eqnarray}\label{teq3}
\nonumber
C_\theta(t) &\simeq& \dfrac{2}{\pi} \left[ 1 + 2\sum_{n=1}^{\infty} \dfrac{(-1)^{n-1}}{4n^2 - 1} \exp(-4n^2 D_r t) \right]
\end{eqnarray}

which captures the general trend of the experimental data (plotted in the inset of Fig. 2c). While the equation looks complex, it can be interpreted as a combination of two separate terms. The second term accounts for the rapid exponential decay due to the rotational diffusion constant, $D_r$, signifying the inherent stochasticity, while the first term of $2/\pi$ quantifies the reorientation condition. The value $2/\pi$ is due to the condition we imposed in our model that the particle's orientation towards home cannot exceed an angle of $\pm\pi/2$. Thus, the above equation captures two important traits of a navigating living agent: its inherent stochasticity and its tolerance for course correction.

Finally, we successfully test our model on a real living system of homing pigeons. We calculated $C_\theta(t)$ for homing pigeon trajectories, as shown in Fig. 2d, and found it exhibits qualitative features similar to those observed in experiments and described by the equation above. Taken together, our results strongly indicate that the paths of homing or migrating animals indeed originate from a combination of two factors: the strength of inherent stochasticity and the frequency of course corrections. Thus, our physical model, inspired by robotic experiments, captures a universal characteristic of homing or migrating trajectories of living organisms.

\section{Summary and outlook}
Here, I introduced a relatively novel field of active and living matter physics. It enables us to understand living systems both as individuals and as collectives, extending traditional condensed matter physics beyond passive, rigid objects. These systems move autonomously using internal energy rather than external fields, thereby rendering equilibrium assumptions and simple Newtonian descriptions ineffective. To this end, I presented simple experimental results from our laboratory on active robots modeled as living organisms. Analyzing the trajectories of homing robots and comparing them with those of real animals reveals universal statistical features in these motions, suggesting that despite biological complexity, the dynamics of living systems display universal statistical behavior.

The examples in this article are drawn mainly from the dynamical behavior of living and active agents, which may create the impression that the scope of this field is relatively narrow. In reality, the field has recently been rebranded as the Physics of Life \cite{PhysicsOfLife} and covers a much broader range of problems within biological sciences. These include how organisms sense their environment and use energy efficiently, how information flows from genes to neurons, the emergence of collective behaviour and pattern formation across length and time scales, evolution, information processing, and the emergence of learning, among many others. The primary focus is on developing predictive physical theories and models that govern this class of problems, so that the field can fully develop into a branch of physics alongside established areas such as astrophysics, condensed matter physics, and particle physics. I will end by providing a few review articles for further reading. \cite{RevModPhysSR,Ramaswamy2010,RevModPhys,Needleman2017ActiveMatter,Volpe_2025}

\section*{Acknowledgements}
The author is grateful for financial support from DST-SERB for CRG Grant No. CRG/2020/002925 and IIT Bombay for the seed grant.

\bibliographystyle{unsrt}
\bibliography{references}

\begin{thebibliography}{1}

\bibitem{PRXLife}
S.~Paramanick, A.~Biswas, H.~Soni, A.~Pal, and N.~Kumar.
\newblock Uncovering universal characteristics of homing paths using foraging
  robots.
\newblock {\em PRX Life}, 2:033007, 2024.

\bibitem{EPJE}
S.~Paramanick, A.~Pal, H.~Soni, and N.~Kumar.
\newblock Programming tunable active dynamics in a self-propelled robot.
\newblock {\em Eur. Phys. J. E}, 47:34, 2024.

\bibitem{PhysicsOfLife}
{National Academies of Sciences, Engineering, and Medicine}.
\newblock {\em Physics of Life}.
\newblock National Academies Press, Washington, DC, 2022.

\bibitem{RevModPhysSR}
M.~C. Marchetti, J.~F. Joanny, S.~Ramaswamy, T.~B. Liverpool, J.~Prost, M.~Rao,
  and R.~A. Simha.
\newblock Hydrodynamics of soft active matter.
\newblock {\em Rev. Mod. Phys.}, 85:1143, 2013.

\bibitem{Ramaswamy2010}
S.~Ramaswamy.
\newblock The mechanics and statistics of active matter.
\newblock {\em Annu. Rev. Condens. Matter Phys.}, 1:323, 2010.

\bibitem{RevModPhys}
S.~R. Nagel.
\newblock Experimental soft-matter science.
\newblock {\em Rev. Mod. Phys.}, 89:025002, 2017.

\bibitem{Needleman2017ActiveMatter}
D.~Needleman and Z.~Dogic.
\newblock Active matter at the interface between materials science and cell
  biology.
\newblock {\em Nat. Rev. Mater.}, 2:17048, 2017.

\bibitem{Volpe_2025}
G.~Volpe, N.~A.~M. Araújo, M.~Guix, M.~Miodownik, N.~Martin, L.~Alvarez,
  J.~Simmchen, R.~Di~Leonardo, N.~Pellicciotta, Q.~Martinet, J.~Palacci, W.~K.
  Ng, D.~Saxena, R.~Sapienza, S.~Nadine, J.~F. Mano, R.~Mahdavi,
  C.~Beck~Adiels, J.~Forth, C.~Santangelo, S.~Palagi, J.~M. Seok, V.~A.
  Webster-Wood, S.~Wang, L.~Yao, A.~Aghakhani, T.~Barois, H.~Kellay,
  C.~Coulais, M.~van Hecke, C.~J. Pierce, T.~Wang, B.~Chong, D.~I. Goldman,
  A.~Reina, V.~Trianni, R.~Beckett, S.~P. Nair, and R.~Armstrong.
\newblock Roadmap for animate matter.
\newblock {\em J. Phys.: Condens. Matter}, 37:333501, 2025.

\end{thebibliography}

\end{document}